\documentclass[%
 preprint,
 showpacs,
 showkeys,
 preprintnumbers,
  amsmath,amssymb,
  aps,
 pra,
  longbibliography,
  floatfix,
 ]{revtex4-1}
\usepackage{hyperref}

\usepackage{graphicx}
\usepackage{epstopdf}
\usepackage{eepic}
\usepackage{xcolor}
\usepackage{braket}
\usepackage{amsmath}
\usepackage{amsthm}

\RequirePackage{times}
\RequirePackage{mathptm}

\newtheorem{Theorem}{Theorem}
\newtheorem{Proposition}[Theorem]{Proposition}

\newtheorem{Fact}[Theorem]{Fact}
\newtheorem{Lemma}[Theorem]{Lemma}
\theoremstyle{definition}
\newtheorem{Definition}[Theorem]{Definition}

\newcommand{\xor}{\text{ ${\tt XOR}$ }}

\begin{document}


\title{A Quantum Random Number Generator Certified by Value Indefiniteness}

\author{Alastair A. Abbott}
\email{aabb009@aucklanduni.ac.nz}
\homepage{http://www.cs.auckland.ac.nz/~aabb009}
\author{Cristian S. Calude}
\email{c.calude@auckland.ac.nz}
\homepage{http://www.cs.auckland.ac.nz/~cristian}
\affiliation{Department of Computer Science, University of Auckland,\\
Private Bag 92019, Auckland, New Zealand}

\author{Karl Svozil}
\email{svozil@tuwien.ac.at}
\homepage{http://tph.tuwien.ac.at/~svozil}
\affiliation{Institut f\"ur Theoretische Physik, Vienna University of Technology,  \\  Wiedner Hauptstra\ss e 8-10/136, A-1040 Vienna, Austria}

\begin{abstract}
In this paper we propose a quantum random number generator (QRNG) which utilizes an entangled photon pair in a Bell singlet state, and is certified explicitly by value indefiniteness. While ``true randomness'' is a mathematical impossibility, the certification by value indefiniteness ensures the quantum random bits are incomputable in the strongest sense. 
  This is the first QRNG setup in which a physical principle (Kochen-Specker value indefiniteness) guarantees that no single quantum bit produced can be classically computed (reproduced and validated), the mathematical form of bitwise physical unpredictability.

The effects of various experimental imperfections are discussed in detail, particularly those related to detector efficiencies, context alignment and temporal correlations between bits. The analysis is to a large extent relevant for the construction of any QRNG based on beam-splitters. By measuring the two entangled photons in maximally misaligned contexts and utilizing the fact that two rather than one bitstring are obtained, more efficient and robust unbiasing techniques can be applied.
A robust and efficient procedure based on ${\tt XOR}$ing the bitstrings together---essentially using one as a one-time-pad for the other---is proposed to extract random bits in the presence of experimental imperfections, as well as a more efficient modification of the von Neumann procedure for the same task. Some open problems are also discussed.
\end{abstract}


\maketitle

\section{Introduction}

Random numbers have been around for more than 4,000 years, but never have they been in such demand as in our time. People use random numbers everywhere.
Thereby, randomness is understood through various ``symptoms.'' Here are three of the largely accepted ones:
\begin{enumerate}
\item[(i)]
{Unpredictability:} It is impossible to win against a random sequence in a fair betting game.
\item[(ii)]
{Incompressibility:}  It is impossible to  compress a random sequence.
\item[(iii)]
{Typicalness:} Random sequences pass every  statistical test of randomness.
\end{enumerate}

Can our intuition on randomness be cast in more rigorous terms? Randomness plays an essential role in probability theory, the mathematical calculus of random events. Kolmogorov axiomatic probability theory assigns probabilities to sets of outcomes and  shows how to calculate with such probabilities;   it assumes randomness, but does not distinguish between individually random and non-random elements.

For example, under a uniform distribution, the outcome of $n$ zeros,
$\underbrace{000\cdots 0}_{n\text{ times}}$,
has the same probability as any other outcome of length $n$, namely $2^{-n}$. A similar situation appears in quantum mechanics: quantum randomness is postulated, not defined or deduced.

Algorithmic information theory (AIT)~\cite{ch6}, developed in the 1960s, defines and studies individual random objects, like finite bitstrings or infinite sequences.
AIT shows that ``pure randomness'' or ``true randomness'' does not exist from a
mathematical point of view.
For example, there is no infinite sequence passing all tests of randomness.
Randomness cannot be mathematically proved: one can never be sure a sequence is random, there are only forms and degrees of randomness.

Computers offer ``random numbers'' produced by algorithms.
Computer scientists needed a long time to realize that randomness produced by software is not random, but only pseudo-random.
This form of randomness mimics well the human perception of randomness, but its quality is rather low because computability destroys many symptoms of randomness,
e.g.\ unpredictability.
It is not totally unreasonable to put forward that pseudo-randomness rather reflects its creators' subjective ``understanding'' and ``projection''  of randomness \footnote{
Psychologists have known for a long time that people tend to distrust streaks in a series of random bits,
hence they imagine a coin flipping sequence alternates between heads and tails much too often for its own sake of ``randomness.''
A simple illustration of this phenomenon, called the gambler's fallacy, is the belief that after a coin has landed on tails ten consecutive times there are more chances that the coin will land on heads at the next flip.}.
And although no computer or software manufacturer claims that their products can generate truly random numbers, recently such formally unfounded claims have re-appeared
for randomness produced with physical experiments
suggesting that ``truly random numbers have been generated at last''~\cite{randomDotOrg,Merali:2014aa}.

\section{Quantum Randomness}

\subsection{Theoretical claims to quantum randomness}

Quantum mechanics has a credible claim to be one of (if not) the best sources of randomness.
There are many quantum phenomena which can be used for random number generation: nuclear decay radiation sources,
the quantum mechanical noise in electronic circuits (known as shot noise), or photons traveling through a semi-transparent mirror.

What is the rationale for the claim that quantum randomness is indeed a better form of randomness than, say, pseudo-randomness?
A quantum random experiment certified by value
indefiniteness---the fact that there can, in general,
be no co- or pre-existing definite values prescribable to certain sets of measurement outcomes~\cite{2008-cal-svo,svozil_2010-pc09}---via the Kochen-Specker Theorem~\cite{kochen1}  generates an {\em infinite (strongly) incomputable sequence of bits}: every Turing machine can reproduce exactly only finitely many scattered digits of such an infinite sequence, i.e.\ the sequence is bi-immune~\cite{2008-cal-svo}.
Such certification, as has already previously been pointed out in~\cite{2008-cal-svo}, is based on the assumption that there are no contextual hidden variables.
Actually, a stronger statement is true:  no Turing machine  can be proved to reproduce exactly any  digit of such an infinite sequence,
i.e.\ it is Solovay bi-immune~\cite{Abbott:aa}.
Indeed, if the value of a bit could be computed before measurement then we could assign a
definite value to the observable, a  contradiction.
The tricky part is that we need to look at infinite sequences to prove the incomputability of individual bits. It is this formal incomputability which corresponds to the physical notion of indeterminism in quantum mechanics---the inability {\em even in principle} to predict the outcome of certain quantum measurements---rather than the mathematically vacuous notion of ``true randomness.''

Quantum random number generators (QRNGs) based on beam splitters~\cite{svozil-qct,rarity-94} have been realized by the Zeilinger group in Innsbruck and Vienna~\cite{zeilinger:qct}
and applied for the sake of violation of {B}ell's inequality under strict {E}instein locality conditions~\cite{zeilinger-epr-98}.

The Gisin group in Geneva~\cite{stefanov-2000}, and in particular its spin-off {\it id Quantique}, produces and markets a commercial device called {\it Quantis}~\cite{Quantis}.
In order to eliminate bias, the device employs von Neumann normalization (actually a more efficient iterated version due to Peres is used~\cite{PeresY-1992}) which requires the {\em independence} of individual events: bits are grouped into pairs, equal pairs (00 or 11) are discarded and we replace 01 with 0 and 10 with 1~\cite{von-neumann1}.

A group in Shanghai and Beijing~\cite{wang:056107} has utilized a Fresnel multiple prism as polarizing beam splitter.
As a normalization technique, previously generated experimental sequences have been used as one time pad to ``encrypt'' random sequences.

QRNGs based on entangled photon pairs have been realized by a second Chinese group in Beijing and Ji'nan~\cite{0256-307X-21-10-027},
who utilized spontaneous parametric down-conversion to produce entangled pairs of photons.
One of the photons has been used as trigger, mostly to allow a faster data production rate by eliminating double counts.
Again, von Neumann normalization has been applied in an attempt to eliminate bias.

A group from the Hewlett-Packard Laboratories in Palo Alto and Bristol~\cite{fiorentino:032334} has used entangled photon pairs
in the Bell basis state $\vert H_1 V_2\rangle + \vert V_1 H_2\rangle$ (note that this is not a singlet state and attains this
form only for one polarization direction; in all the other directions the state contains also $V_1V_2$ as well as $H_1H_2$ contributions), where the outcomes $H_1,V_1$ and $H_2,V_2$
refer to observables associated with unspecified (presumably identical for both particles) directions.
In analogy to von Neumann normalization, the coincidence events $H_1V_2$ and  $V_1H_2$ have been mapped into 0  and 1, respectively.
Thereby, as the authors have argued, the 2-qubit space of the photon pair is effectively restricted
to a two-dimensional Hilbert subspace described by an effective-qubit state.

A more recent rendition of a QRNG~\cite{10.1038/nature09008}, although not based on photons and beamsplitters, utilizes Boole-Bell-type setups ``secured by''
Boole-Bell-type inequality violations
in the spirit of quantum cryptographic protocols~\cite{ekert91,PhysRevLett.85.3313}.
This provides some indirect ``statistical verification'' of value indefiniteness (again under the assumption of noncontextuality),
but falls short of providing certification of strong incomputability
{\it via} value indefiniteness~\cite{2008-cal-svo,svozil-2009-howto}.
With regard to value indefiniteness,
the difference between Boole-Bell-type inequalities  {\it versus} Kochen-Specker-type theorems is this:
In the Boole-Bell-type case,
the breach of value indefiniteness needs not happen at every single particle,
whereas in the Kochen-Specker-type case this must happen {\em for every particle}~\cite{svozil_2010-pc09}.
Pointedly stated, the Boole-Bell-type violation is statistical, but {\em not necessarily} on every quantum separately.
Hence, because a Boole-Bell-type violation does not guarantee that every bit is certified by value indefiniteness,
one could potentially produce sequences containing infinite computable subsequences
``protected'' by Boole-Bell-type violations.
Further,   given that such criticisms seem also to hold for the statistical verification of value indefiniteness~\cite{panbdwz,huang-2003,cabello:210401}, it seems unlikely that statistical tests of the measurement outcomes alone can fully certify such a QRNG.
\if01
Further, given the ability to simulate maximal violation of Bell-type inequalities with Mealy automata~\cite{Cabello:PC2010}, such ``certification'' fails to truly certify the incomputability of the QRNG; this incomputability is the formal manifestation of quantum indeterminism, and it is this which the QRNG must be certified for. Indeed, given that such criticisms seem also to hold for the statistical verification of value indefiniteness~\cite{panbdwz,huang-2003,cabello:210401}, it seems unlikely that statistical tests of the measurement outcomes alone can fully certify such a QRNG.
\fi
\if01
(ii)   \marginpar{\small OK?}
A device operating essentially in four-dimensional Hilbert space is protected by quantum value indefiniteness in the
strongest Kochen-Specker-type form
(subject to the validity of quantum theory and the willingness to accept proofs by contradiction
as a valid physical argument).
Alas, no proof exists that the binary sequences obtained when a measurement device or a subsequent algorithm,
instead of the quaternary observable,
produces {\em dichotomic} observables based on the former quaternary
ones,
inherits this strong form of value indefiniteness from the quaternary case.
\fi

\subsection{Shortcomings of current QRNGs}

It is clear that any QRNG claiming a better quality of randomness has to produce at least an infinite incomputable sequence of outputs, preferably  a strongly incomputable one.
Do the current proposals of QRNGs generate ``in principle'' strongly incomputable sequences of quantum random bits?
To answer this question one has to check whether the QRNG is ``protected'' by value indefiniteness, the only physical principle currently known to guarantee incomputability; in most cases the answer is either negative or cannot be verified because of lack of information about the mechanism of the QRNG.

In Ref.~\cite{PhysRevA.82.022102} tests based on algorithmic information theory were used to analyze and compare quantum and non-quantum bitstrings.
Ten strings of length $2^{32}$ bits each from two quantum sources (the commercial {\em Quantis} device \cite{idQuantique} and the Vienna Institute for Quantum Optics and Quantum Information group~\cite{ThomasJennewein}) and three classical sources (Mathematica, Maple and the binary expansion of $\pi$) were analyzed.
No distribution was assumed for any of the sources, yet a test based on Borel-normality was able to distinguish between the quantum and non-quantum sources of random numbers. It is known that all algorithmically random strings are Borel-normal \cite{calude:02}, although the converse is not true. Indeed, the tests found the quantum sources to be less normal than the pseudo-random  ones. Is this a property of quantum randomness, or evidence of flaws in the tested QRNGs?

In Ref.~\cite{AbbottCalude10} the probability distribution for an ideal QRNG was discussed: not surprisingly, such devices are seen to sample from the uniform distribution.
Testing the same strings as in~\cite{PhysRevA.82.022102} against this expected distribution, strong evidence was found that the QRNGs tested are {\em not} sampling from the correct distribution. Further, weaker evidence suggests the pseudo-random sources of randomness---Mathematica and Maple---are, on the contrary, too normal. The results of the analysis are presented in Table~\ref{pvalues}.


\begin{table}
\begin{center}
\begin{tabular}{cccccc}
\hline\hline
QRNG &  $k=1$ & $k=2$ & $k=3$ & $k=4$ & $k=5$\\
\hline
{\bf Maple} & 0.79 & 0.15 & 0.83 & 0.47 & 0.97\\
{\bf Mathematica} & 0.18 & 0.38 & 0.35 & 0.45 & 0.99\\
{\bf $\pi$} & 0.38 & 0.27 & 0.05 & 0.62 & 0.21\\
{\bf Quantis} & $\mathbf{< 10^{-10}}$ & $\mathbf{< 10^{-10}}$ & $\mathbf{< 10^{-10}}$ & $\mathbf{< 10^{-10}}$ & $\mathbf{< 10^{-10}}$\\
{\bf Vienna} & 0.12 & $\mathbf{< 10^{-10}}$ & $\mathbf{< 10^{-10}}$ & $\mathbf{< 10^{-10}}$ & $\mathbf{< 10^{-10}}$\\
\hline\hline
\end{tabular}
\end{center}
\caption{$p$-values for the $\chi^2$ test that the bitstring is sampled from the uniform distribution. Bold values indicate statistically significant evidence that the strings are not sampled from the uniform distribution.}
\label{pvalues}
\end{table}

The notable exception to these findings are the Vienna bits which, when viewed at the single-bit level, appear unbiased.
It appears that the good performance at the 1-bit level has been achieved (perhaps through experimental feedback control) at the sacrifice of the performance at the $k\ge 2$ level, a property much harder to control without post-processing.
The {\it Quantis} QRNG uses iterated von Neumann normalization in an attempt to unbias the output; the fact that this is not completely successful indicates either a significant variation in bias over time, or non-independence of successive bits~\cite{AbbottCalude10}.

These results highlight the need to pay extra attention in the design process to the distribution produced by a QRNG. Normalization techniques are an effective way to remove bias, but to have the desired effect assumptions about independence and constancy of bias must be satisfied~\cite{AbbottCalude10}. While experiments will never realize the ideal QRNG, one needs to be aware of how much affect experimental imperfections have.
Any credible QRNG should take these issues into account, as well as the need of explicit certification of randomness  by some physical law, e.g.\ value indefiniteness.

\section{The scheme under ideal conditions}

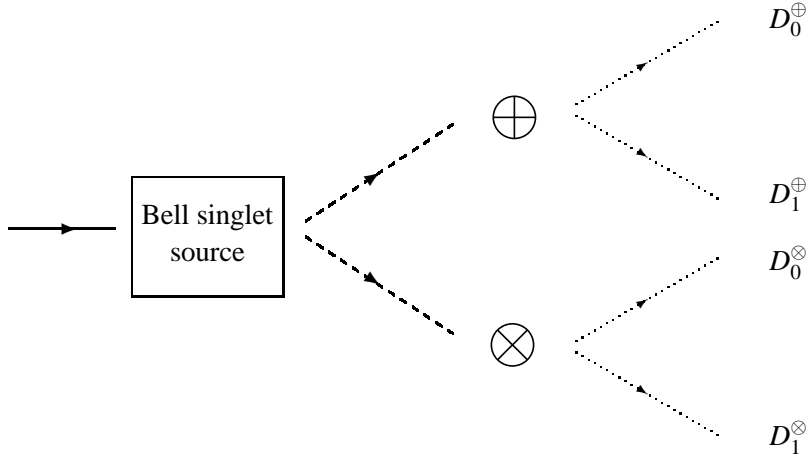
\begin{figure}
\begin{center}

\unitlength 1mm 
\linethickness{0.4pt}
\ifx\plotpoint\undefined\newsavebox{\plotpoint}\fi 
\begin{picture}(104,55.75)(0,0)
\put(16.75,19){\framebox(20,15.75)[cc]{}}
\put(26.75,29.5){\makebox(0,0)[cc]{Bell singlet}}
\put(26.75,24.5){\makebox(0,0)[cc]{source}}
\thicklines
\put(9.375,28){\vector(1,0){.07}}\put(.25,28){\line(1,0){14.25}}
\put(49.625,35.5){\vector(3,2){.07}}\multiput(39.68,28.93)(.049375,.0325){16}{\line(1,0){.049375}}
\multiput(41.26,29.97)(.049375,.0325){16}{\line(1,0){.049375}}
\multiput(42.84,31.01)(.049375,.0325){16}{\line(1,0){.049375}}
\multiput(44.42,32.05)(.049375,.0325){16}{\line(1,0){.049375}}
\multiput(46,33.09)(.049375,.0325){16}{\line(1,0){.049375}}
\multiput(47.58,34.13)(.049375,.0325){16}{\line(1,0){.049375}}
\multiput(49.16,35.17)(.049375,.0325){16}{\line(1,0){.049375}}
\multiput(50.74,36.21)(.049375,.0325){16}{\line(1,0){.049375}}
\multiput(52.32,37.25)(.049375,.0325){16}{\line(1,0){.049375}}
\multiput(53.9,38.29)(.049375,.0325){16}{\line(1,0){.049375}}
\multiput(55.48,39.33)(.049375,.0325){16}{\line(1,0){.049375}}
\multiput(57.06,40.37)(.049375,.0325){16}{\line(1,0){.049375}}
\multiput(58.64,41.41)(.049375,.0325){16}{\line(1,0){.049375}}
\put(49.625,20.5){\vector(3,-2){.07}}\multiput(39.68,26.93)(.049375,-.0325){16}{\line(1,0){.049375}}
\multiput(41.26,25.89)(.049375,-.0325){16}{\line(1,0){.049375}}
\multiput(42.84,24.85)(.049375,-.0325){16}{\line(1,0){.049375}}
\multiput(44.42,23.81)(.049375,-.0325){16}{\line(1,0){.049375}}
\multiput(46,22.77)(.049375,-.0325){16}{\line(1,0){.049375}}
\multiput(47.58,21.73)(.049375,-.0325){16}{\line(1,0){.049375}}
\multiput(49.16,20.69)(.049375,-.0325){16}{\line(1,0){.049375}}
\multiput(50.74,19.65)(.049375,-.0325){16}{\line(1,0){.049375}}
\multiput(52.32,18.61)(.049375,-.0325){16}{\line(1,0){.049375}}
\multiput(53.9,17.57)(.049375,-.0325){16}{\line(1,0){.049375}}
\multiput(55.48,16.53)(.049375,-.0325){16}{\line(1,0){.049375}}
\multiput(57.06,15.49)(.049375,-.0325){16}{\line(1,0){.049375}}
\multiput(58.64,14.45)(.049375,-.0325){16}{\line(1,0){.049375}}
\put(67.5,42.75){\makebox(0,0)[cc]{{\Huge $\oplus$}}}
\put(67.25,12.5){\makebox(0,0)[cc]{{\Huge $\otimes$}}}
\thinlines
\put(85.125,50){\vector(3,2){.07}}\multiput(75.68,44.43)(.81522,.47826){24}{{\rule{.4pt}{.4pt}}}
\put(85.125,6){\vector(3,-2){.07}}\multiput(75.68,11.43)(.81522,-.47826){24}{{\rule{.4pt}{.4pt}}}
\put(85.125,37.5){\vector(3,-2){.07}}\multiput(75.68,42.93)(.81522,-.47826){24}{{\rule{.4pt}{.4pt}}}
\put(85.125,18.5){\vector(3,2){.07}}\multiput(75.68,12.93)(.81522,.47826){24}{{\rule{.4pt}{.4pt}}}
\put(104,55.75){\makebox(0,0)[cc]{$D^\oplus_0$}}
\put(104,32.5){\makebox(0,0)[cc]{$D^\oplus_1$}}
\put(104,23.5){\makebox(0,0)[cc]{$D^\otimes_0$}}
\put(104,.25){\makebox(0,0)[cc]{$D^\otimes_1$}}
\end{picture}
\end{center}
\caption{Scheme of a quantum random number generator~\cite{svozil-2009-howto}.}
\label{2010-qxor-f1}
\end{figure}

In what follows, a proposal for a QRNG depicted in Fig.~\ref{2010-qxor-f1}, previously put forward in Ref.~\cite{svozil-2009-howto}, will be discussed in detail.
It utilizes the singlet state of two two-state particles (e.g., photons of linear polarization) proportional to  $\vert H_1 V_2\rangle - \vert V_1 H_2\rangle$, which is form invariant in all measurement directions.

A single photon light source (presumably an LED) is attenuated so more than one photons are rarely in the beam path at the same time.
These photons impinge on a source of singlet states of photons (presumably by spontaneous parametric down-conversion in a nonlinear medium).
The two resulting entangled photons are then analyzed with respect to their linear polarization state at some directions which are $\pi /4$ radians  ``apart,''
symbolized by ``$\oplus$'' and ``$\otimes$,''
respectively.

Due to the required four-dimensional Hilbert space, this  QRNG
is ``protected'' by Bell- as well as Kochen-Specker- and Greenberger-Horne-Zeilinger-type value
indefiniteness \footnote{Note that this is not the case for current QRNGs based on beam-splitters,
which operate in a Hilbert space of dimension two.}.
The protocol utilizes all three principal types of quantum indeterminism:
(i) the indeterminacy of individual outcomes of single events as proposed by Born and Dirac;
(ii)  quantum complementarity (due to the use of conjugate variables), as put forward by Heisenberg, Pauli and Bohr; and
(iii) value indefiniteness due to Bell, Kochen \& Specker, and Greenberger, Horne \& Zeilinger.

This, essentially, is the same experimental configuration as the one used for a measurement of the correlation function at the angle of $\pi /4$ radians ($45^\circ$).
Whereas the correlation function averages over ``a large number'' of single contributions, a random sequence can be obtained by concatenating these single pairs of outcomes via addition modulo~2.

Formally, suppose that for the $i$th experimental run, the two outcomes are
$O^\oplus_i \in \{0,1\}$ corresponding to $D^\oplus_0$ or  $D^\oplus_1$,
and
$O^\otimes_i \in \{0,1\}$  corresponding to $D^\otimes_0$ or  $D^\otimes_1$.
These two outcomes $O^\oplus_i $ and  $O^\otimes_i$, which themselves form two sequences of random bits,
are subsequently combined by the ${\tt XOR}$ operation, which amounts to their parity, or to the addition modulo 2 according to Table~\ref{2010-qxor-t1} (in what follows, depending on the formal context,
 ${\tt XOR}$ refers to either a binary function
of two binary observables, or to the logical operation).
Stated differently, one outcome is used as a {\em one time pad} to ``encrypt'' the other outcome,
and {\it vice versa}.
\begin{table}
\begin{center}
\begin{tabular}{cccccc}
\hline\hline
$O^\oplus_i $ &  $O^\otimes_i$ & $O^\oplus_i $ ${\tt XOR}$  $O^\otimes_i$\\
\hline
0 & 0 & 0 \\
0 & 1 & 1 \\
1 & 0 & 1 \\
1 & 1 & 0 \\
\hline\hline
\end{tabular}
\end{center}
\caption{The logical exclusive or operation.}
\label{2010-qxor-t1}
\end{table}
As a result, one obtains a sequence $x=x_1x_2\ldots x_n$ with
\begin{equation}
x_i=O^\oplus_i + O^\otimes _{i} \text{ mod }2 .
\label{2010-qxor-e1}
\end{equation}

For the ${\tt XOR}$d sequence to still be certifiably incomputable (via value indefiniteness), one must prove this certification is preserved under ${\tt XOR}$ing---indeed strong incomputability itself is {\em not} necessarily preserved. By necessity any QRNG certified by value indefiniteness must operate non-trivially in a Hilbert space of dimension $n\ge 3$. To transform the $n$-ary (incomputable) sequence into a binary one, a function $f: \{0,1,\dots, n-1\} \to \{0,1,\lambda\}$ must be used ($\lambda$ is the empty string); to claim certification, the strong incomputability of the bits must still be guaranteed after the application of $f$. This is a fundamental issue which has to be checked for  existing QRNGs such as that in Ref.~\cite{10.1038/nature09008}; without it one cannot claim to produce truly indeterministic bits. In general incomputability itself is not preserved by $f$; however by consideration of the value indefiniteness of the source the certification can be seen to hold under ${\tt XOR}$ as well as when discarding bits~\cite{Abbott:aa}.
\if01
In order to be certified by the strongest form of Kochen-Specker-type quantum value indefiniteness,
the device has to be constantly monitored
for such strong violations of value definiteness \cite{panbdwz,huang-2003,cabello:210401}.
(Note that, as has been argued earlier, due to the statistical character of Boole-Bell-type violations of value definiteness,
monitoring for violations of Boole-Bell type inequalities is insufficient.)
Note also that it it is still an open question wether or not the ``degradation''
of $n$-ary to two outcomes preserves value indefiniteness.
One strategy might be to {\em assume} (without proof)
that, for $n>2$,   the $n$-ary sequence $s=s_1,s_2,s_3,\cdots $ based on $n$-ary outcomes is asymptotically random.
Then, the binary sequence $s'=s_1',s_2',s_3',\cdots $ with $s_i\in \{0,1,\ldots, n-1\}$
obtained through the omission of $n-2$ symbols

\begin{equation*}
s_i' =
	\begin{cases}
		s_i, & \mbox{if } s_i\in \{0,1\},\\
		\emptyset, & \mbox{else }
	\end{cases}
\end{equation*}
remains asymptotically random \cite{calude:02}.
 \marginpar{\small OK?}
\fi

\section{``Random'' errors or systematic errors}

In what follows we shall discuss possible ``random'' (no pun) or systematic errors in experimental realizations of this QRNG (many of these errors may appear in other types of photon-based QRNGs.)  Our aim is to draw attention to the specific nature of such errors and how they affect the resulting bitstrings. A good QRNG must, in addition to the necessary certification (e.g.\ by value indefiniteness), take into account the nature of these errors and be carefully designed (along with any subsequent post-processing) so that the resultant distribution of bitstrings the QRNG samples from is as close as possible to the expected uniform distribution~\cite{AbbottCalude10}. Both the uniformity of the source and incomputability are ``independent symptoms'' of randomness, and care must be taken to obtain both properties.

\subsection{Double counting}

One conceivable problem is that
the detectors analyzing the different polarization directions do not respond to photons of the same pair,
but to two photons belonging to different pairs.
This seems to be no drawback for the application of the ${\tt XOR}$ operation since
(at least in the absence of temporal correlations between bits)
the postulates of quantum mechanics state that
the individual outcomes occur independently and indeterministically (the last property is mathematically  modeled by strong incomputability~\cite{2008-cal-svo,Abbott:aa}).
If, however, events are not independent then more care is needed. However, correlation between events is an undesirable property in itself, and as long as care is made, it is unlikely to be made worse by double counting.

\subsection{Non-singlet states}

The state produced by the  spontaneous parametric down-conversion may not be exactly a singlet.
This may give rise to a systematic bias of the combined light source-analyzer setup in a very similar way as for beam splitters.

\subsection{Non-alignment of polarization measurement angles}\label{sec:nonalignment}

No experimental realization will attain a ``perfect anti-alignment'' of the polarization analyzers  at angles $\pi /4$ radians apart.
Only in this ideal case are the bases conjugate and the correlation function will be exactly zero.
Indeed, ``tuning'' the angle to obtain equi-balanced sequences of zeroes and ones may be a method to properly anti-align the polarizers.
However, one has to keep in mind that any such ``tampering'' with the raw sequence of data
to achieve  Borel normality (e.g.\ by readjustments of the experimental setup)
may introduce unwanted (temporal) correlations or other bias~\cite{PhysRevA.82.022102}.

Incidentally, the angle $\pi /4$ is one of the three points at angles $0$, $\pi /4$ and $\pi /2$ in the interval $[0, \pi /2]$
in which the classical and quantum correlation functions coincide.
For all other angles, there is a higher ratio of different or identical pairs than could be expected classically.
Thus, ideally, the QRNG could be said to operate in the ``quasi classical'' regime,
albeit fully certified by quantum value indefiniteness.

Quantitatively, the expectation function of the sum of the two outcomes modulus~2 can be defined by  averaging over the sum modulo~2 of the outcomes $O^0_i, O^\theta _i\in \{0,1\}$ at angle $\theta$ ``apart''
in the $i$th experiment, over a ``large number'' of experiments; i.e.,
\begin{equation*}
E_{\tt XOR}(\theta )=\lim_{N \rightarrow \infty} {1\over N}\sum_{i=1}^N \left( O^0_i + O^\theta _i \text{ mod }2\right).
\end{equation*}
This is related to the standard correlation function,
\begin{equation*}
	C(\theta)=\lim_{N\to \infty}{1\over N}\sum_{i=1}^N O^0_i \cdot O^\theta_i
\end{equation*}
by
\begin{equation*}
	E_{\tt XOR}(\theta)=\frac{|C(\theta)-1|}{2} \raisebox{0.5ex}{,}
\end{equation*}
where
\begin{equation*}
	O^0_i \cdot O^\theta_i =
	\begin{cases}
		1, & \mbox{if }O^0_i = O^\theta_i,\\
		-1, & \mbox{if }O^0_i \neq O^\theta_i.
	\end{cases}
\end{equation*}
A detailed calculation yields the classical linear expectation function
$E^{\text{cl}} _{\tt XOR}(\theta ) = {1-2 \theta / \pi}$,
and the quantum expectation function
$E_{\tt XOR}(\theta ) = (1/2)(1+\cos 2\theta )$.
\begin{figure}[h]
\begin{center}
	\includegraphics[scale=0.7]{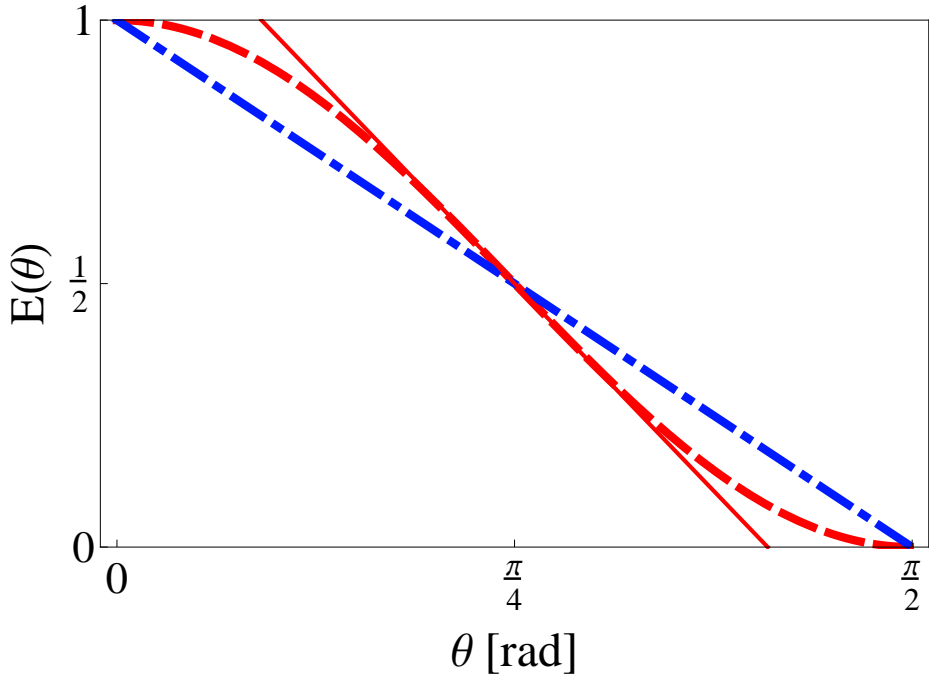}
\end{center}
\caption{(Color online) The classical and quantum expectation functions and the linear quantum approximation around $\pi/4$.}
\label{corrFuncs}
\end{figure}

Thus, for angles ``far apart'' from $\pi /4$, the ${\tt XOR}$ operation actually {\em deteriorates} the two random signals taken from the two analyzers {\em separately.}
The deterioration is even {\em greater quantum mechanically than classically,} as the entangled particles are more correlated and thus ``less independent.''
Potentially, this could be utilized to ensure a $\pi/4$ mismatch more accurately than possible through classical means.
This will be discussed in section~\ref{ss-orthogonality} below.

In order to avoid this negative feature while generating bits, instead of ${\tt XOR}$ing outcomes of {\em identical} partner pairs, one could ${\tt XOR}$  time-shifted outcomes; e.g., instead of the expression in Eq.~(\ref{2010-qxor-e1})
one may consider
\begin{equation}
x_i=O^0_i + O^\theta _{i+j} \text{ mod }2, \text{ with } j> 0.
\label{2010-qxor-e2}
\end{equation}
One should make $j$ large enough so that, taking in to account double counting, there is no chance of accidentally causing two offset but correlated outcomes to be ${\tt XOR}$'d together. Theoretical analysis of the effects of experimental imperfections and the ${\tt XOR}$ operation are discussed later in the paper, and ${\tt XOR}$ing shifted pairs is an efficient and effective procedure for reducing such errors.

\subsection{Different detector efficiencies}

Differences in detector efficiencies result in a bias of the sequence. This  complicating effect is separate from non-perfect misalignment of polarization context.
Suppose that the probabilities of detection are denoted by
$p_{H_1}$,
$p_{H_2}$,
$p_{V_1}$,
$p_{V_2}$. Since  $p_{H_1}+p_{V_1}=p_{H_2}+p_{V_2}=1$,
the probability to find pairs adding up to 0 and 1 modulo~2 are
$p_{H_1}p_{H_2}+ p_{V_1}p_{V_2}=1 - (p_{H_1}+p_{H_2}) + 2p_{H_1}p_{H_2}$
and
$p_{H_1}p_{V_2}+ p_{V_1}p_{H_2}=p_{H_1}+p_{H_2} - 2p_{H_1}p_{H_2}$, respectively (adding up to 1). If both $p_{H_1}\neq p_{V_1}$ and $p_{H_2}\neq p_{V_2}$ then the resulting ${\tt XOR}$'d sequence is biased. The two obtained sequences could be unbiased before or after ${\tt XOR}$ing by the von Neuman method \cite[p. 768]{von-neumann1}, although any temporal correlations would violate the condition of independence required by this method. One should keep in mind, however, that the von Neumann normalization procedure necessarily discards many bits (more efficient methods exist~\cite{PeresY-1992}). The efficiency can be increased by utilizing both strings more carefully, and such a method is discussed in Section~\ref{sec:crits}.

\subsection{Unstable detector bias}

Von Neumann type normalization procedures will only remove bias due to detector efficiencies if the bias remains constant over time. If the bias drifts over time due to instability in the detectors, the resulting normalized sequence will not be unbiased but instead will simply be less biased~\cite{AbbottCalude10}. It is difficult to overcome this, as experimental instability is inevitable. However, bounds on the bias of the normalized sequence based on reasonable experimental parameters~\cite{AbbottCalude10} can be used to determine the length for which the source samples ``closely enough'' from the uniform distribution.

If the bias varies independently between detectors, the ${\tt XOR}$ing process should serve to reduce the impact of varying detector efficiencies and applying von Neumann normalization to the ${\tt XOR}$'d bitstring is advantageous compared working with a single bitstring from a source of varying bias.

\subsection{Temporal correlations, photon clustering and ``bunching''}

Due to the Hanbury-Brown-Twiss effect, the photons may be temporally correlated and thus arrive clustered or ``bunched.''
Temporal correlations appear also at ``double-slit analogous experiments'' in the time domain~\cite{PhysRevLett.95.040401},
in which the role of the slits is played by windows in time of attosecond duration.  This can, to an extent, be avoided by ensuring successive photons are sufficiently separated, although this poses a limit on the bitrate of such a device. However, since the case where two or more singlet pairs are in the beam path at once is potentially of sufficient importance, this effect needs further careful consideration.

Another conceivable source of temporal correlations is due to the detector dead-time, $T_d$, during which the detector is inactive after measurement~\cite{stefanov-2000}. If we measure $O^\oplus_i=0$, the detector $D^\oplus_0$ corresponding to $0$ is unable to detect another photon for a small amount of time, significantly increasing the chance of detecting a photon at the other detector during this time, obtaining a $1$. This leads to higher than expected chances of $01$ and $10$ being measured. This is problematic as such a correlation will not be removed by ${\tt XOR}$ing, even with an offset of $j$. However, this can be avoided by discarding any measurements within time $T_d$ from the previous measurement.

In view of conceivable temporal correlations, it would be interesting to test the quality of the random signal as $j$ is varied in Eq.~(\ref{2010-qxor-e2}). As previously mentioned, any temporal correlations  will violate the condition of independence needed for von Neumann normalization making it difficult to remove any bias in the output;  if the dependence can be bounded then unbiasing techniques such as that proposed by Blum~\cite{18499} could be used instead of von Neumann's procedure. It seems desirable and simpler to avoid temporal correlations with carefully designed experimental methodology as opposed to post-processing where possible.

\subsection{Fair sampling}

As in most optical tests of Bell's inequalities ~\cite{clauser,chau},
the inefficiency of photon detection requires us to make the {\em fair sampling assumption}~\cite{PhysRevD.35.3831,PhysRevA.57.3304,Pearle:1970fk,PhysRevA.81.012109}:  the loss is independent of the measurement settings, so the ensemble of detected systems provides a fair statistical sample of the total ensemble. In other words, we must exclude the possibility of a ``demon'' in the measuring device conspiring against us in choosing which bits to reject.

The strength of the proposed QRNG relies crucially on value indefiniteness,
 so without this fair sampling assumption we would forfeit the assurance of bitwise incomputability of the generated sequence.
As an example let us consider the extreme case that the detection efficiency is less that 50\%;
our supposed demon could reject all bits detected as 0 and be within the bounds given by this efficiency,
while the produced sequence would be computable. In the more general case for any  efficiency $\rho < 1$
 the demon could reject bits to ensure every $\left(1/(1-\rho)\right)$'th bit is a zero;
 this would introduce an infinite computable subsequence, a property violating the strong incomputability of the output bitstring
 produced by our QRNG,
 and still be consistent with the detection efficiency.

Note that this condition is stronger than the fair sampling assumption required in tests for violation of Bell-type inequalities because, without this assumption,  {\it any} inefficiency can lead to a loss of randomness.

\section{Better-than-classical operationalization of spatial orthogonality}
\label{ss-orthogonality}

As has already been pointed out, for no temporal offset and
in the regime of relative spatial angles around $\pi/4$ --- i.e., at almost half orthogonal measurement directions ---
the classical linear expectation function
$E^{\text{cl}} _{\tt XOR}(\theta ) = {1-2 \theta / \pi}$,
for $0 < \theta < \pi /4$
is strictly {\em smaller},
and for $\pi /4 < \theta < \pi /2$ is strictly {\em greater}
than the quantum expectation function
$E_{\tt XOR}(\theta ) = (1/2)(1+\cos 2\theta )$.
This can be demonstrated by rewriting
$\theta =\pi /4 \pm \Delta \theta$,
and by considering  a Taylor series expansion around $\pi /4$
for small $\Delta \theta  \ll 1$, which yields
$E_{\tt XOR}(\pi/4\pm \Delta \theta  ) \approx (1/2) \mp  \Delta \theta $,
whereas
$E^{\text{cl}} _{\tt XOR}(\pi/4\pm \Delta \theta ) = (1/2) \mp (2 / \pi )\Delta \theta $ (see Fig.~\ref{corrFuncs}).

Phenomenologically this indicates
less-than-classical numbers of equal pairs of outcomes ``0--0'' as well as   ``1--1,''
and
more-than-classical non-equal pairs of outcomes ``0--1'' as well as   ``1--0,'' respectively,
for the quantum case in the region $0 < \theta < \pi /4$; as well as the reverse behavior in the region $\pi /4 < \theta < \pi /2$.
This  in turn results in ``less zeroes'' and ``more ones'' of the resulting sequence obtained by ${\tt XOR}$ing the pairs of outcomes in the region $0 < \theta < \pi /4$,
as well as in ``more zeroes'' and ``less ones'' in the region $\pi /4 < \theta < \pi /2$ as compared to classical non-entangled systems~\cite{peres222}.
Hence, with increasing aberration from misalignment  $\Delta \theta$ the quantum device ``drifts off'' into biasedness of the output  ``faster'' than any classical device.
As a result, Borel normality is expected to be broken more strongly and quickly quantum mechanically than classically.

This effect could in principle be used to operationalize spatial orthogonality through the fine-tuning of angular directions yielding Borel normality.
In the resulting protocols, quantum mechanics outperforms any classical scheme due to the differences in the correlation functions.

\section{Theoretical analysis on generated bitstrings}

Here we analyze the output distribution of the proposed QRNG and the ability to extract uniformly distributed bits from the two generated bitstrings in the presence of experimental imperfections.

\subsection{Probability space construction}

With reference to Fig.~\ref{2010-qxor-f1} for the setup,
we write the generated Bell singlet state with respect the top (``$\oplus$'') measurement context
(this is arbitrary as the singlet is form invariant in all measurement directions) as
$\frac{1}{\sqrt{2}}(\ket{01} - \ket{10})$.
 The lower (``$\otimes$'') polarizer is at an angle of $\theta$ to the top one. After beam splitters we have the state
$$\frac{1}{\sqrt{2}}\left[\cos \theta(\ket{00}-\ket{11}) - \sin \theta (\ket{01} + \ket{10})\right],$$
 so we measure the same outcome in both contexts with probability $\cos^2\theta$ and different outcomes with probability $\sin^2\theta$.

\if01
\begin{Fact}\label{fact:TrigBinom}
	For any $k > 0$, $0\le \theta < 2\pi$ and $x\in B^k$ we have $$\sum_{y\in B^k}(\sin^2 \theta)^{d(x,y)}(\cos^2\theta)^{k-d(x,y)} = 1,$$ where $d(x,y)$ is the Hamming distance, i.e. the number of places at which $x$ and $y$ differ.
\end{Fact}
\begin{proof}
	We have
	\begin{align*}
		\sum_{y\in B^k}(\sin^2 \theta)^{d(x,y)}(\cos^2\theta)^{k-d(x,y)} &= \sum_{i=0}^k\binom{k}{i}(\sin^2 \theta)^{i}(\cos^2\theta)^{k-i}\\
		&= (\cos^2\theta + \sin^2\theta)^k\\
		&= 1.
	\end{align*}
\end{proof}
\begin{Proposition}\label{outputProbSpace}
	The probability space produced by the QRNG is $(B^n\times B^n,2^{B^n\times B^n},P_{n^2})$ where $P_{n^2}: 2^{B^n\times B^n} \to [0,1]$ is defined for all $X\subseteq B^n\times B^n$ as:
	$$P_{n^2}(X)=\sum_{(x,y)\in X}2^{-n}(\sin^2\theta)^{d(x,y)}(\cos^2\theta)^{n-d(x,y)},$$
	where $d(x,y)$ is the standard Hamming distance.
\end{Proposition}
We check easily that this is indeed a valid probability space (i.e.\ that is satisfies the Kolmogorov axioms~\cite{Billingsley:1979aa}).
\begin{enumerate}
	\item $P_{n^2}(\emptyset) = 0$, trivially true;
	\item $P_{n^2}(B^n\times B^n) = 2^{-n}\sum_{(x,y)\in B^n\times B^n}(\sin^2\theta)^{d(x,y)}(\cos^2\theta)^{n-d(x,y)} = 2^{-n} 2^n = 1$ by Fact~\ref{fact:TrigBinom};
	\item For $X,Y\subseteq B^n\times B^n$, $X\cap Y=\emptyset \implies P_{n^2}(X \cup Y) = P_{n^2}(X) + P_{n^2}(Y)$, trivially true.
\end{enumerate}

\begin{Fact}
	For any $x\in B^n$ we have
	\begin{align*}
		P_{n^2}(\{x\} \times B^n) &= P_{n^2}(B^n \times \{x\})\\
		&= 2^{-n}\sum_{k=0}^n\binom{n}{k}(\sin^2\theta)^k(\cos^2\theta)^{n-k}\\
		&= 2^{-n}.
	\end{align*}
\end{Fact}
So we see that each string taken separately is uniformly distributed; if we discard one string and use the QRNG to generate one bit string only then the output will be uniformly distributed.

\subsection{Independence of the QRNG probability space}
If we were to discard one string it is clear the other string is produced independently (since it is uniformly distributed). However, we would like to extend our notion of independence defined in~\cite{AbbottCalude10} to this 2-string probability space.

\begin{Definition}\label{defn:independence}
	We say the probability space $(B^n\times B^n,2^{B^n\times B^n},R_{n^2})$ is independent if for all $1\le k \le n$ and $x_1,\dots,x_k$, $y_1,\dots,y_k \in B^k$ we have
	\begin{align*}
		R_{n^2}(x_1\dots x_k B^{n-k} \times y_1\dots y_k B^{n-k})=&R_{n^2}(x_1\dots x_{k-1} B^{n-k+1} \times y_1\dots y_{k-1} B^{n-k+1})\\
		&\times R_{n^2}(B^{k-1}x_k B^{n-k} \times B^{k-1} y_k B^{n-k}).
	\end{align*}
\end{Definition}

\begin{Lemma}\label{independenceLemma}
	For all $x,y \in B^{|x|}$ and $0\le k + |x| \le n$ we have $$R_{n^2}(B^{n-k}xB^{n-k-|x|} \times B^{n-k}yB^{n-k-|x|}) = 2^{-|x|}(\sin^2\theta)^{d(x,y)}(\cos^2\theta)^{|x|-d(x,y)}.$$
\end{Lemma}
\begin{proof}
	We have
	\begin{align*}
		R_{n^2}(B^{n-k}xB^{n-k-|x|} \times B^{n-k}yB^{n-k-|x|}) =& \sum_{a_1,a_2 \in B^{n-k}}\sum_{b_1,b_2\in B^{n-k-|x|}}R_{n^2}\left((a_1 x b_1,a_2 y b_2) \right)\\
		=& 2^{-n}(\sin^2\theta)^{d(x,y)}(\cos^2\theta)^{|x|-d(x,y)}\\&\cdot\sum_{a_1,a_2,b_1,b_2}(\sin^2\theta)^{d(a_1b_1,a_2b_2)}(\cos^2\theta)^{n-|x| -d(a_1b_1,a_2b_2)}\\
		=& 2^{-n}(\sin^2\theta)^{d(x,y)}(\cos^2\theta)^{|x|-d(x,y)}2^{n-|x|}\\
		=& 2^{-|x|}(\sin^2\theta)^{d(x,y)}(\cos^2\theta)^{|x|-d(x,y)}
	\end{align*}
\end{proof}

\begin{Fact}
	The probability space $P_{n^2}$  defined in Proposition~\ref{outputProbSpace} is independent.
\end{Fact}
\begin{proof}
	This follows from Lemma~\ref{independenceLemma} and the fact that the Hamming distance is additive, i.e.\ $d(x_1\dots x_k, y_1\dots y_k)=d(x_1\dots x_{k-1},y_1\dots y_{k-1})+d(x_k,y_k)$.
\end{proof}

\subsection{XOR application}

We now consider the situation where the two output strings $x$ and $y$ are XOR'd against each other (effectively using one as a one-time pad for the other) to produce a single string, and we investigate the distribution of this resulting string.

For $j\ge 0$ and $x,y \in B^{n+j}$ define $X_j: B^{n+j} \times B^{n+j} \to B^n$ as $X_j(x,y) = z$ where $z_i = x_i \oplus y_{i+j}$ for $i=1,\dots,n$. For $z \in B^n$ let
\begin{align*}
	A_j(z) &= \{(x,y) \mid x,y \in B^{n+j}, X_j(x,y)=z \}\\
	&= \{(ua,b(u \text{ \tt XOR } z) \mid u \in B^n, a,b\in B^j \}.
\end{align*}
\begin{Proposition}
 The probability space of interest is $(B^n,2^{B^n},Q_{n,j})$, where $Q_{n,j}: 2^{B^n} \to [0,1]$ is defined for all $X\subseteq B^n$ as:
\begin{align*}
	Q_{n,j}(X) =& \sum_{z\in X}P_{(n+j)^2}(A(z)).
\end{align*}
\end{Proposition}
We note that $|A(z)| = 2^{n+2j}$ and check this is a valid probability space:
\begin{enumerate}
	\item $Q_{n,j}(\emptyset) = 0$, trivially true.
	\item $Q_{n,j}(B^n) = \sum_{z\in B^n}P_{(n+j)^2}(A(z)) = P_{(n+j)^2}(\bigcup_z A(z)) = P_{(n+j)^2}(B^{n+j}\times B^{n+j}) = 1$, where we have noted that all $A(z)$ are disjoint and thus $|\bigcup_z A(z)| = 2^n 2^{n+2j} = (2^{n+j})^2$ and hence $\bigcup_z A(z) = B^{n+j}\times B^{n+j}$.
	\item For $X,Y \subseteq B^n$, $X\cap Y = \emptyset \implies Q_{n,j}(X\cup Y) = Q_{n,j}(X) + Q_{n,j}(Y)$, trivially true.
\end{enumerate}

\begin{Theorem}
	For $j>0$, $Q_{n,j}=U_n$, the uniform distribution, and for $j=0$, $Q_{n,j} = P_n$, the constantly biased distribution from the von Neumann note with $p_0 = \cos^2\theta$ and $p_1 = 1-p_0 = \sin^2\theta$.
\end{Theorem}
\begin{proof}
	Let $z \in B^n$ and $j \ge 0$. Then we have
	\begin{align*}
		Q(z) =& P_{n^2}(A(z))\\
		=& \sum_{a,b\in 2^j}\sum_{u\in 2^n}P_{n^2}((ua,b(u \text{ \tt xor } z))\\
		=& 2^{-n-j}\sum_{u\in 2^n}(\sin^2\theta)^{d(u[j+1,n],(u\text{ \tt xor } z)[1,n-j])}(\cos^2\theta)^{n-j-d(u[j+1,n],(u\text{ \tt xor } z)[1,n-j])}\\&\cdot \sum_{a\in 2^j}(\sin^2\theta)^{d(a,(u\text{ \tt xor } z)[n-j+1,n])}(\cos^2\theta)^{j-d(a,(u\text{ \tt xor } z)[n-j+1,n])} \\&\cdot\sum_{b\in 2^j}(\sin^2\theta)^{d(u[1,j],b)}(\cos^2\theta)^{j-d(u[1,j],b)}\\
		=& 2^{-n-j}\sum_{u\in 2^n}(\sin^2\theta)^{d(u[j+1,n],(u\text{ \tt xor } z)[1,n-j])}(\cos^2\theta)^{n-j-d(u[j+1,n],(u\text{ \tt xor } z)[1,n-j])},
	\end{align*}
	where the last line follows from Fact~\ref{fact:TrigBinom}. Now note that for $j=0$, $d(u,(u\text{ \tt xor } z)) = \#_1(z)$. Hence for $j=0$ we have
\begin{align*}
	Q(z) &= 2^{-n}\sum_{u\in 2^n}(\sin^2\theta)^{\#_1(z)}(\cos^2\theta)^{\#_0(z)}\\
	&= (\sin^2\theta)^{\#_1(z)}(\cos^2\theta)^{\#_0(z)}.
\end{align*}

For $j>0$ we now have
\begin{align*}
	Q(z) =& 2^{-n-j}\sum_{u_n\in B}\cdots \sum_{u_{n-j}\in B}(\sin^2\theta)^{u_n \oplus z_{n-j}\oplus u_{n-j}}(\cos^2\theta)^{1-u_n \oplus z_{n-j}\oplus u_{n-j}}\cdots\\&\times \sum_{u_1 \in B} (\sin^2\theta)^{u_j+1 \oplus z_{1}\oplus u_{1}}(\cos^2\theta)^{1-u_j+1 \oplus z_{1}\oplus u_{1}}\\
	=& 2^{-n-j}\sum_{u_n\in B}\cdots \sum_{u_{n-j}\in B}(\sin^2\theta + \cos^2\theta)\cdot \sum_{u_1 \in B}(\sin^2\theta + \cos^2\theta)\\
	=& 2^{-n-j}\sum_{u_{n-j+1}\dots u_n \in B^j}1\\
	=& 2^{-n-j}2^j\\
	=& 2^{-n}.
\end{align*}
Hence for $j>0$  the resulting probability space is uniformly distributed.
\end{proof}
\fi


More formally, the QRNG generates two strings simultaneously, so the probability space contains pairs of strings of length $n$. Let $e_x^\oplus,e_y^\otimes$ for $x,y=0,1$ be the detector efficiencies of the $D_x^\oplus$ and $D_y^\otimes$ detectors respectively. For perfect detectors, i.e\ $e_x^\oplus = e_y^\otimes$, we would expect a pair of bits $(a,b)$ to be measured with probability $2^{-1} (\sin^2\theta)^{a \oplus b}(\cos^2\theta)^{1-a \oplus b}$; non-perfect detectors alter this probability depending on the values of $a,b$.

Let $B=\{0,1\}$, and for $x,y \in B^n$ let $d(x,y)$ be the Hamming distance between the strings $x$ and $y$, i.e\ the number of positions at which $x$ and $y$ differ, and let $\#_b(x)$ be the number of $b$s in $x$.

	The probability space \footnote{$B^{n}$ is the set of bitstrings  $x$ of length $|x|=n$; $2^{X}$ is the set of all subsets of the set $X$.} of bitstrings produced by the QRNG is $(B^n\times B^n,2^{B^n\times B^n},P_{n^2})$, where the probability $P_{n^2}: 2^{B^n\times B^n} \to [0,1]$ is defined for all $X\subseteq B^n\times B^n$ as follows:
	$$P_{n^2}(X)=\frac{1}{Z_n}\sum_{(x,y)\in X}(\sin^2\theta)^{d(x,y)}(\cos^2\theta)^{n-d(x,y)}(e_0^\oplus)^{\#_0(x)}(e_1^\oplus)^{\#_1(x)}(e_0^\otimes)^{\#_0(y)}(e_1^\otimes)^{\#_1(y)},$$
	
\noindent	and the term
	\begin{align*}
		Z_n&=\sum_{(x,y)\in B^n\times B^n}(\sin^2\theta)^{d(x,y)}(\cos^2\theta)^{n-d(x,y)}(e_0^\oplus)^{\#_0(x)}(e_1^\oplus)^{\#_1(x)}(e_0^\otimes)^{\#_0(y)}(e_1^\otimes)^{\#_1(y)}\\
		&= \left[(\sin^2\theta(e_0^\oplus e_1^\otimes + e_1^\oplus e_0^\otimes)+\cos^2\theta(e_0^\oplus e_0^\otimes + e_1^\oplus e_1^\otimes)  \right]^n
	\end{align*}
ensures normalization.

We can check easily that this is indeed a valid probability space (i.e.\ that is satisfies the Kolmogorov axioms~\cite{Billingsley:1979aa}).
\if01
\begin{enumerate}
	\item $P_{n^2}(\emptyset) = 0$, trivially true;
	\item $P_{n^2}(B^n\times B^n) = 1$ by by definition of $Z_n$;
	\item For $X,Y\subseteq B^n\times B^n$, $X\cap Y=\emptyset \implies P_{n^2}(X \cup Y) = P_{n^2}(X) + P_{n^2}(Y)$, trivially true.
\end{enumerate}
\fi
Note that for equal detector efficiencies we have
\begin{align*}
	Z_n&=(e^\oplus)^n(e^\otimes)^n\sum_{(x,y)\in B^n\times B^n}(\sin^2\theta)^{d(x,y)}(\cos^2\theta)^{n-d(x,y)}=2^n(e^\oplus)^n(e^\otimes)^n,
\end{align*}
hence the probability has the simplified form
$$P_{n^2}(X)=\sum_{(x,y)\in X}2^{-n}(\sin^2\theta)^{d(x,y)}(\cos^2\theta)^{n-d(x,y)}.$$

Given that the proposed QRNG produces two (potentially correlated) strings, it is worth considering the distribution of each string taken separately. Given the rotational invariance of the singlet state this should be uniformly distributed. However, because the detector efficiencies may vary in each detector, this is not, in general, the case.
	For every bitstring $x\in B^n$ we have
	\begin{align}
		P_{n^2}(\{x\} \times B^n) &= \frac{1}{Z_n}\sum_{y\in B^n}(\sin^2\theta)^{d(x,y)}(\cos^2\theta)^{n-d(x,y)}(e_0^\oplus)^{\#_0(x)}(e_1^\oplus)^{\#_1(x)}(e_0^\otimes)^{\#_0(y)}(e_1^\otimes)^{\#_1(y)}\notag\\
		&= \frac{(e_0^\oplus)^{\#_0(x)}(e_1^\oplus)^{\#_1(x)}}{Z_n}\sum_{y\in B^n}(\sin^2\theta)^{d(x,y)}(\cos^2\theta)^{n-d(x,y)}(e_0^\otimes)^{\#_0(y)}(e_1^\otimes)^{\#_1(y)}\notag\\
		&= \frac{1}{Z_n}\left(e_0^\oplus (e_1^\otimes\sin^2\theta+e_0^\otimes\cos^2\theta) \right)^{\#_0(x)}\left( e_1^\oplus (e_0^\otimes\sin^2\theta+e_1^\otimes\cos^2\theta) \right)^{\#_1(x)}.\label{discardOneStrDis}
	\end{align}

We see that each bitstring taken separately appears to come from a constantly biased source where the probabilities that a bit is 0 or 1, $p_0,p_1$, are given by the formulae
$$p_0 = e_0^\oplus (e_1^\otimes\sin^2\theta+e_0^\otimes\cos^2\theta)/Z_1, \, p_1 = e_1^\oplus (e_0^\otimes\sin^2\theta+e_1^\otimes\cos^2\theta)/Z_1.$$
This can alternatively be viewed as the distribution obtained if we were to discard one bitstring after measurement.
Note that if either $e_0^\otimes = e_1^\otimes$ or we have perfect misalignment (i.e.\ $\theta=\pi/4$) then the probabilities have the simpler formulae: $$p_x=e_x^\oplus/(e_0^\oplus + e_1^\oplus), x\in\{0,1\}.$$ In this case, if we further have that $e_0^\oplus = e_1^\oplus$, we obtain the uniform distribution by discarding one string after measurement.

The analogous result for the symmetrical case $P_{n^2}\left(B^n \times \{y\} \right)$ also holds.

\subsection{Independence of the QRNG probability space}
If we were to discard one bitstring it is clear the other bitstring is generated independently in a statistical sense since the probability distribution source producing it is  constantly biased and independent~\cite{AbbottCalude10}. However, we would like to extend our notion of independence defined in~\cite{AbbottCalude10} to this 2-bitstring probability space.

	We say the probability space $(B^n\times B^n,2^{B^n\times B^n},R_{n^2})$ is {\it independent} if for all $1\le k \le n$ and $x_1,\dots,x_k$, $y_1,\dots,y_k \in B$ we have
	\begin{align*}
		R_{n^2}(x_1\dots x_k B^{n-k} \times y_1\dots y_k B^{n-k})=& \, R_{n^2}(x_1\dots x_{k-1} B^{n-k+1} \times y_1\dots y_{k-1} B^{n-k+1})\\
		&\times R_{n^2}(B^{k-1}x_k B^{n-k} \times B^{k-1} y_k B^{n-k}).
	\end{align*}

	For all $x,y \in B^{|x|}$ and $0\le k + |x| \le n$ we have
	$$P_{n^2}(B^{n-k}xB^{n-k-|x|} \times B^{n-k}yB^{n-k-|x|}) = P_{|x|^2}((x,y)).$$
	Indeed, using the additivity of the Hamming distance and the $\#_x$ functions, e.g. $d(x_1\dots x_k, y_1\dots y_k)=d(x_1\dots x_{k-1},y_1\dots y_{k-1})+d(x_k,y_k)$, we have:
	\begin{align*}
		P_{n^2}(B^{n-k}xB^{n-k-|x|} \times B^{n-k}yB^{n-k-|x|}) =& \sum_{a_1,a_2 \in B^{n-k}}\sum_{b_1,b_2\in B^{n-k-|x|}}P_{n^2}\left((a_1 x b_1,a_2 y b_2) \right)\\
		=& P_{|x|^2}((x,y))\sum_{a_1,a_2 \in B^{n-k}}\sum_{b_1,b_2\in B^{n-k-|x|}}P_{(n-|x|)^2}\left((a_1 b_1,a_2 b_2) \right)\\
		=& P_{|x|^2}((x,y)) P_{(n-|x|)^2}(B^{n-|x|}\times B^{n-|x|})\\
		=& P_{|x|^2}((x,y)).
	\end{align*}


	As a direct consequence we deduce that the probability space $P_{n^2}$  defined above
	is independent.

\subsection{XOR application}

We now consider the situation where the two output bitstrings $x$ and $y$ are ${\tt XOR}$'d against each other (effectively using one as a one-time pad for the other) to produce a single bitstring, and we investigate the distribution of the resulting bitstring. Rather than only considering the effect of ${\tt XOR}$ing paired (and potentially correlated) bits, we also consider ${\tt XOR}$ing outcomes shifted by $j>0$ bits as described in Section~\ref{sec:nonalignment}.

For $j\ge 0$ and $x,y \in B^{n+j}$ define the offset-${\tt XOR}$ fucntion $X_j: B^{n+j} \times B^{n+j} \to B^n$ as $X_j(x,y) = z$ where $z_i = x_i \oplus y_{i+j}$ for $i=1,\dots,n$. For $z \in B^n$ the set of pairs $(x,y)$ which produce $z$ when ${\tt XOR}$'d with offset $j$ is
\[	A_j(z) = \{(x,y) \mid x,y \in B^{n+j}, X_j(x,y)=z \}
	= \{(ua,b(u \xor z) \mid u \in B^n, a,b\in B^j \}.\]
 The probability space of the output produced by the QRNG is $(B^n,2^{B^n},Q_{n,j})$, where $Q_{n,j}: 2^{B^n} \to [0,1]$ is defined for all $X\subseteq B^n$ as:
\begin{align}\label{QProbDefn}
	Q_{n,j}(X) =& \sum_{z\in X}P_{(n+j)^2}(A_j(z)).
\end{align}

We note that $|A_j(z)| = 2^{n+2j}$ and check this is a valid probability space. Indeed,
	 $Q_{n,j}(\emptyset) = 0$, is trivially true,
	$$Q_{n,j}(B^n) = \sum_{z\in B^n}P_{(n+j)^2}(A_j(z)) = P_{(n+j)^2}\left(\bigcup_z A_j(z)\right) = P_{(n+j)^2}\left(B^{n+j}\times B^{n+j}\right) = 1,$$ bcause all $A_j(z)$ are disjoint and thus $$|\bigcup_z A_j(z)| = 2^n 2^{n+2j} = (2^{n+j})^2, \mbox{  so  }  \bigcup_z A_j(z) = B^{n+j}\times B^{n+j},$$ and
	for disjoint  $X,Y \subseteq B^n$ we have $ Q_{n,j}(X\cup Y) = Q_{n,j}(X) + Q_{n,j}(Y)$.

We now explore the form of the ${\tt XOR}$'d distribution $Q_{n,j}$  for  $j=0$ and $j>0$.

	Let $z \in B^n$ and $j \ge 0$. By $z[m,k]$ we denote the substring $z_m\dots z_k, 1\le m \le k\le n$. We have
	\begin{align*}
		Q_{n,j}(z) =& P_{(n+j)^2}(A_j(z)))\\
		=& \sum_{a,b\in 2^j}\sum_{u\in 2^n}P_{(n+j)^2}((ua,b(u\xor z))\\
		=& \sum_{u\in 2^n}P_{(n-j)^2}\left((u[j+1,n],(u\xor z)[1,n-j])\right)  \\
		&\cdot \sum_{a\in 2^j}P_{j^2}\left((a,(u\xor z)[n-j+1,n])\right)  \sum_{b\in 2^j}P_{j^2}\left((u[1,j],b)\right).
	\end{align*}
	For $j=0$, we note that $d(u, u\xor z) = \#_1(z)$, and thus we have:
\begin{align*}
	Q_{n,0}(z) &= \sum_{u\in 2^n}P_{n^2}\left((u,(u\xor z))\right) \\
		&= \frac{1}{Z_n}(\sin^2\theta)^{\#_1(z)}(\cos^2\theta)^{\#_0(z)} \sum_{u\in B^n} (e_0^\oplus)^{\#_0(u)}(e_1^\oplus)^{\#_1(u)}(e_0^\otimes)^{\#_0(u\xor z)}(e_1^\otimes)^{\#_1(u\xor z)}\\
		&= \frac{1}{Z_n}\left(\sin^2\theta (e_0^\oplus e_1^\otimes + e_1^\oplus e_0^\otimes) \right)^{\#_1(z)}\left(\cos^2\theta (e_0^\oplus e_0^\otimes + e_1^\oplus e_1^\otimes) \right)^{\#_0(z)}.
\end{align*}
We recognize this as a constantly biased source where $$p_0 = \cos^2\theta (e_0^\oplus e_0^\otimes + e_1^\oplus e_1^\otimes)/Z_1,\, p_1 = \sin^2\theta (e_0^\oplus e_1^\otimes + e_1^\oplus e_0^\otimes)/Z_1.$$ It is interesting to compare the form of $Q_{n,0}$ to the distribution of the constantly biased source Eq.~\eqref{discardOneStrDis} by discarding one output string---the former is more sensitive to misalignment, the latter to differences in detection efficiencies. In the case of perfect/equal detector efficiencies (but non-perfect misalignment), discarding one string produces uniformly distributed bitstrings, whereas ${\tt XOR}$ing does not.

We now look at the case where $j>0$. For the ideal situation of $\theta=\pi/4$ we have the same result as for the $j=0$ case, while if we have equal detector efficiencies then we get the uniform distribution. We show this as follows (note that $Z_{n+j}=2^{n+j}$ in this case):
\begin{align*}
	Q_{n,j}(z) =& 2^{-n-j}\sum_{u_n\in B}\cdots \sum_{u_{n-j}\in B}(\sin^2\theta)^{u_n \oplus z_{n-j}\oplus u_{n-j}}(\cos^2\theta)^{1-u_n \oplus z_{n-j}\oplus u_{n-j}}\cdots\\&\times \sum_{u_1 \in B} (\sin^2\theta)^{u_j+1 \oplus z_{1}\oplus u_{1}}(\cos^2\theta)^{1-u_j+1 \oplus z_{1}\oplus u_{1}}\\
	=& 2^{-n-j}\sum_{u_n\in B}\cdots \sum_{u_{n-j}\in B}(\sin^2\theta + \cos^2\theta)\cdot \sum_{u_1 \in B}(\sin^2\theta + \cos^2\theta)\\
	=& 2^{-n-j}\sum_{u_{n-j+1}\dots u_n \in B^j}1\\
	=& 2^{-n}.
\end{align*}
However, in the more general case of non-equal detector efficiencies, the distribution is no longer independent, although in general is much closer to the uniform distribution than the $j=0$ case.
(Recall that independence is a sufficient but not necessary condition for uniform distribution~\cite{AbbottCalude10}.) It is indeed this ``closeness''---the total variation distance given by $\Delta(U_n,Q_{n,j}) = \frac{1}{2}\sum_{x\in B^n}|2^{-n} - Q_{n,j}(x)|$---which is the important quantity ($U_n$ is the uniform distribution on $n$-bit strings). However, since $Q_{n,j}$ for $j>0$ is not independent, von Neumann normalization cannot be applied to guarantee the uniform distribution; indeed the dependence is not even bounded to a fixed number of preceding bits.
\begin{table}
\begin{center}
\begin{tabular}{cccc}
\hline\hline
$x$ & $\text{bin}(174)$ & $\text{bin}(487)$ & $\text{bin}(973)$\\
\hline
$Q_{10,0}(x)$ & $5.90\times 10^{-4}$ & $9.70\times 10^{-4}$ & $1.64\times 10^{-4}$\\
$Q_{10,1}(x)$ & $9.75\times 10^{-4}$ & $9.71\times 10^{-4}$ & $9.71\times 10^{-4}$\\
$Q_{10,2}(x)$ & $9.78\times 10^{-4}$ & $9.70\times 10^{-4}$ & $9.70\times 10^{-4}$\\
$U_{10}(x)$ & $9.77\times 10^{-4}$   & $9.77\times 10^{-4}$ & $9.77\times 10^{-4}$\\
\hline\hline
\end{tabular}
\end{center}
\caption{Emperical evidence for the quality of ${\tt XOR}$ing with $j>0$ compared to $j=0$ and configuration settings of $\theta=\pi/5$, $e_0^\oplus=0.30$, $e_1^\oplus=0.33$, $e_0^\otimes=0.29$, $e_1^\otimes=0.30$ --- this is probably much worse (further from the ideal case) that one would expect in an experimental setup. The (small) value of $n=10$ has been used as, unfortunately, the distribution is very costly to calculate numerically. Here $\text{bin}(m)$ denotes the (10-bit zero-extended) binary representation of $m$. For example, $\text{bin}(1)=0000000001$, $\text{bin}(2)=0000000010$, etc. }
\label{SampleQValues}
\end{table}
\begin{table}
\begin{center}
\begin{tabular}{cccc}
\hline\hline
$\Delta(Q_{10,0},U_{10})$ & 0.770271 &&\\
$\Delta(Q_{10,1},U_{10})$ & \phantom{0.} 0.00441399 &&\\
$\Delta(Q_{10,1},U_{10})$ & \phantom{0.} 0.00440061 &&\\
\hline\hline
\end{tabular}
\end{center}
\caption{The variation from the uniform distribution of the distributions $Q_{10,j}$, using the same parameters as Table~\ref{SampleQValues}.}
\label{QDistEx}
\end{table}
\begin{figure}[ht]
\begin{center}
\includegraphics[scale=1.08]{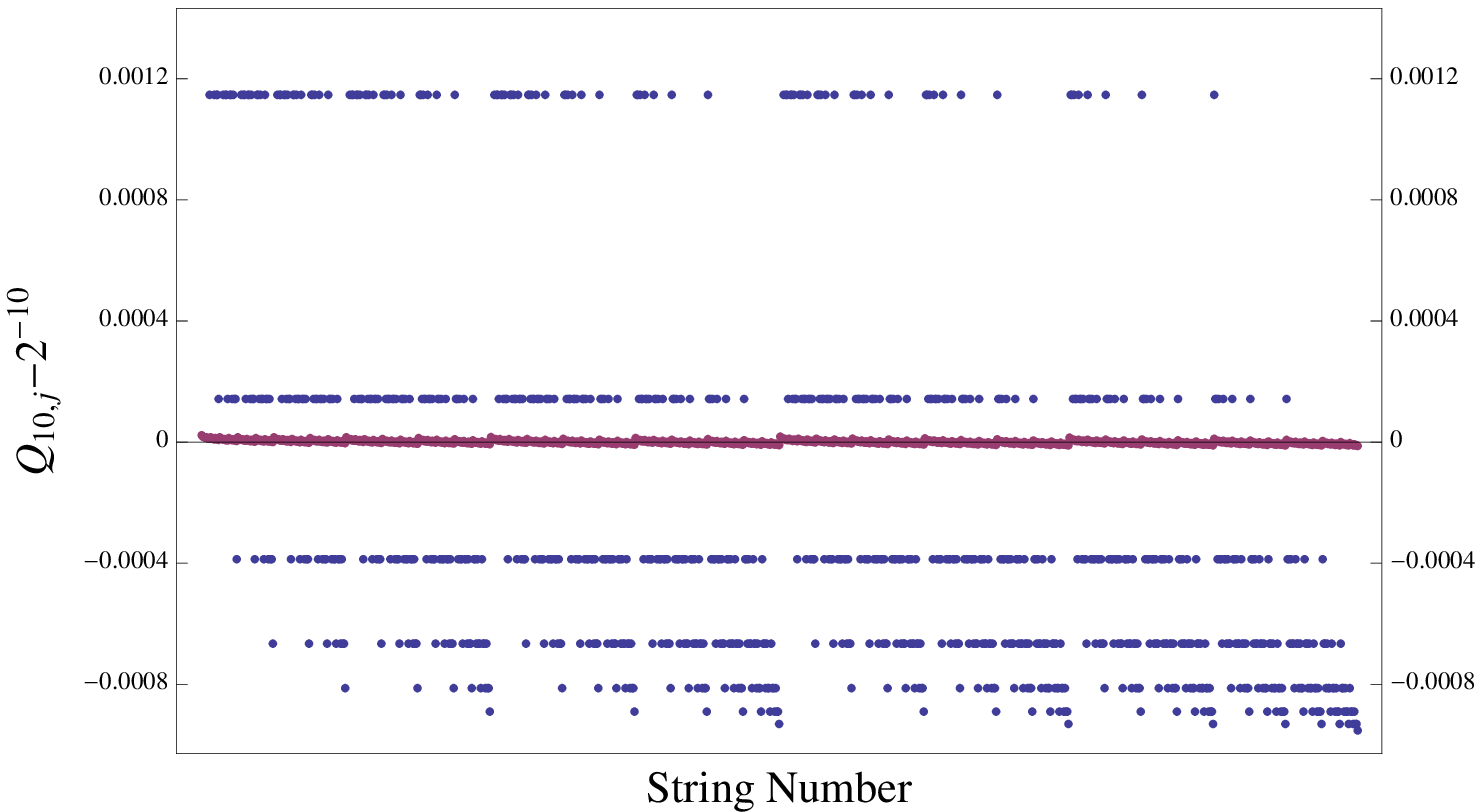}
\end{center}
\caption{(Color online) A plot of $Q_{10,j}-2^{-10}$ for each of the $2^{10}$ strings of length 10. The two cases $j=0$ (blue) and $j=1$ (red) show how much closer the probabilities given by $Q_{10,1}$ are to that expected from the uniform distribution than for $Q_{10,0}$. The same experimental configuration as in Table~\ref{QDistEx} has been used.}
\label{VarFromUD}
\end{figure}

\subsection{Criticisms and alternative operationalizations}\label{sec:crits}

This given, one may ask why not simply discard one string to give the distribution in Eq.~\eqref{discardOneStrDis} and apply von Neumann normalization to obtain uniformly distributed bitstrings. There are two primary answers to this question.

(i) As discussed previously the effect of drift in bias and temporal correlations will ensure this method will not produce the uniform distribution anyway. Indeed, the distribution $Q_{n,j}$ for $j>0$ should be more robust to those effects ($Q_{n,j}$ for example is less sensitive to detector bias than that in Eq.~\eqref{discardOneStrDis}). It is extremely plausible that $Q_{n,j}$ gives as good results as discarding one string in practice; it is indeed very close to the uniform distribution as can be seen from Table~\ref{QDistEx} and Fig.~\ref{VarFromUD}. To compare properly the distributions, the following \emph{open question} must be answered: what is the bound $\rho$ depending on $e_x^\oplus,e_y^\otimes$ and $\theta$ such that $\Delta(U_n,Q_{n,j})\le \rho$, and how does that compare to that given in~\cite{AbbottCalude10} for normalization of a source with varying bias?

Further, $Q_{n,j}$ produces bitstrings of length $n$, whereas applying von Neumann to a single string produces a string with expected length at most $n/4$ bits.
This is a significant increase in efficiency, making the shifted ${\tt XOR}$ing process extremely appealing for a high bitrate, un-normalized QRNG.
Even the $j=0$ case with von Neumann applied after ${\tt XOR}$ing would often be preferable to discarding one string, since it is less sensitive to detector efficiency (the hardware limit) and more sensitive to to misalignment (which is controlled by the experimenter).

(ii) If one insists on a perfect theoretical distribution in the presence of non-ideal misalignment and unequal detector efficiencies, or perhaps the $Q_{n,j}$ distribution is not sufficient for particular requirements, then one can still operationalize both strings to improve the efficiency of the QRNG over discarding a single string by a simple modification of von Neumann's procedure. To do so, note that the pair of pairs $(a_1 a_2, b_1 b_2)$ have the same probability as the pairs $(a_2 a_1, b_2 b_1)$. By mapping those with $a_1 b_1 < a_2 b_2$ (lexicographically)  to $0$, those with $a_1 b_1 > a_2 b_2$ to 1, and discarding those with $a_1 b_1 = a_2 b_2$, one will obtain the uniform distribution as for von Neumann's procedure. The key advantage is that this will obtain strings of expected length up to $3n/8$, while maintaining the desired property of sampling from the  uniform distribution.

The problem of determining how best to obtain the maximum amount of information from the QRNG is largely a problem of randomness extractors~\cite{Gabizon:2010uq}, and is a trade off between the number of uniformly distributed bits obtained and the processing cost---a suitable extractor needs to operate in real-time for most purposes. As we have seen, the fact that two (potentially correlated) bitstrings are obtained allows more efficient operation than a QRNG using single-photons. We have shown how the proposed QRNG can be operationalized in more than one way: either by using shifted ${\tt XOR}$ing of bits to sample from a distribution which is close to (equal to in the ideal limit) the uniform distribution and efficient and robust to various errors, or by utilizing both produced bitstrings to allow a more efficient normalization procedure giving (in absence of the aforementioned temporal effects) the uniform distribution. Many more operationalizations are undoubtedly possible.

\section{Summary}

Every QRNG  claiming to produce a better form of randomness than pseudo-randomness must firstly be certified by some physical law implying the incomputability of the output bitstrings; value indefiniteness is one such example. Most existing proposals of QRNGs are based on single beam splitters and work in a dimension-two Hilbert space, so they cannot be certified by value indefiniteness given by  the Kochen-Specker theorem (which holds only in a Hilbert space of dimension greater than 2).
In this paper we have proposed a QRNG which, by utilizing an entangled photon singlet-state in four-dimensional Hilbert space, is certified by value indefiniteness which implies strong incomputability, the mathematical property corresponding to physical indeterminism. While this is an ingredient of fundamental importance in any reasonable QRNG, we have recognized that experimental imperfections will always prevent the QRNG from producing exactly the theoretical uniform probability distribution, another essential symptom of randomness (independent of incomputability). The form and effects of these conceivable experimental errors have been discussed, and care has been taken to make the proposed QRNG robust to these effects.

Since this QRNG produces two bitstrings, we have proposed ${\tt XOR}$ing the bitstrings produced---using one as a one-time pad for the other---to obtain better protection against experimental imperfections, particuarly non-ideal misalignment and unequal detector efficiencies, and utilize the benefit of these two strings over simply using one. Rather than ${\tt XOR}$ing corresponding bits, bits $x_i$ and $y_{i+j}$ are ${\tt XOR}$'d (for fixed $j>0$) as this not only provides much better results, but also mitigates the effects of temporal correlations between adjacent bits. Further, we have proposed an alternative normalization method based on von Neumann's procedure which uses both bitstrings. This procedure is significantly more efficient yet still guarantees uniformly distributed strings in the presence of non-ideal misalignment and unequal detector efficiencies.
We leave it as an \emph{open question} to improve upon the time-shifted ${\tt XOR}$ method and find a technique to extract bits which are provably uniformly distributed  and is more efficient than the improved von Neumann method discussed.

Analyses of sequences generated by the proposed QRNG should be conducted, utilizing the knowledge of the expected uniform distribution, as in~\cite{PhysRevA.82.022102}. In particular, the quality of both the individual strings produced should be compared with that of the ${\tt XOR}$'d sequence, both with and without von Neumann normalization applied, as well as the sequence produced by our improved von Neumann method.

Further, in view of conceivable temporal correlations between bits, the quality of the random bits should be tested as $j$ is varied in Eq.~\eqref{QProbDefn}. Since this has little effect on the bias of the resultant string (and normalization can subsequently remove this), it would allow investigation of the effect and significance of these conceivable temporal correlations.

The proposed QRNG produces bits which are both certified via value indefiniteness and should be distributed more uniformly than those produced by existing QRNGs based on beam splitters. It will be interesting to experimentally test the quality of bits produced via this method against existing classical and quantum sources of randomness.

\section*{Acknowledgment} We thank A. Cabello and A. Zeilinger for many interesting discussions about quantum randomness.


%

\end{document}